# Illustrating the space-time coordinates of the events associated with the apparent and the actual position of a light source

Bernhard Rothenstein[1)], Stefan Popescu[2)] and George J. Spix[3)]

1) Politehnica University of Timisoara, Physics Department, Timisoara, Romania
2) Siemens AG, Erlangen, Germany
3) BSEE Illinois Institute of Technology, USA

*Abstract. We present a space- time diagram that displays in true values the space- time coordinates of events associated with the apparent and actual positions of a point like source moving with constant velocity. We use it in order to construct the actual shape of a moving luminous profile or to determine the relationship between apparent and actual distances. We show that the simple fact that light propagates at finite speed has important consequences like "length contraction" or "length dilation" the effects being sensitive against the "approaching" and the "receding" character of the source relative to an observer located at the origin of the inertial reference frame.*

## 1. Introduction

Deissler[1] presents some physical effects associated with the fact that light propagates with finite speed $c$. The effects are detected from the rest frame K(XOY) of an observer $R_0(0,0)$ equipped with a clock $T_0(0,0)$ located at the origin $O$. At each point of the space defined by the axes of the K reference frame we find an observer $R(x,y)$ with his clock $T(x,y)$. All the clocks display the same running time as a result of a clock synchronization procedure proposed by Einstein.

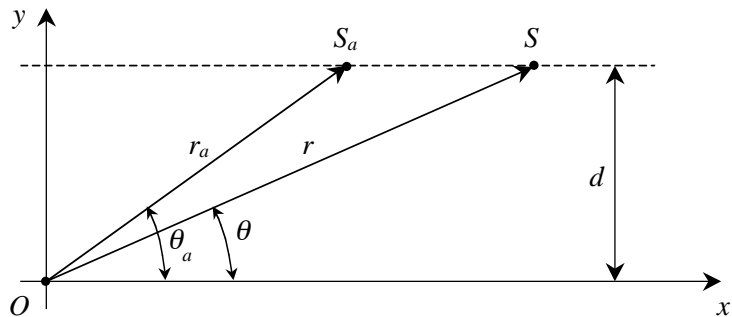

**Figure 1.** *Apparent position $S_a$ and actual position $S$ of a source of light that moves with constant velocity $V = \beta \cdot c$ parallel to the OX axis at a distance d apart of it. The two positions are defined by the space coordinates $S_a(x_a = r_a \cos\theta_a, y_a = r_a \sin\theta_a)$ and $S(x = r\cos\theta, y = r\sin\theta)$ respectively, using polar and Cartesian coordinates as well.*

In Figure 1 $S_a(x_a, y_a) = S_a(r_a \cos\theta_a, r_a \sin\theta_a)$ represents a position of the point like source. The source moves with constant velocity $V$ parallel to and in the positive direction of the OX axis at a distance *d* apart of it. A clock $T_a(x_a, y_a)$ located at this point reads $t_a = -\frac{r_a}{c}$ when the source S emits a light signal that will be received by $R_0(0,0)$ when his clock $T_0(0,0)$ reads a zero time. $S_a$ represents the *apparent position* of source S. On the same figure $S(x, y) = S(r\cos\theta, r\sin\theta)$ represents the *actual position* of the source, located at this point when the observer $R_0(0,0)$ receives the light signal that was previously emitted from the apparent position $S_a$. At the same moment the clock $T(x, y) = T(r\cos\theta, r\sin\theta)$ located at the actual position reads a zero time too. When $R_0(0,0)$ receives a second light signal emitted from the actual position S his clock $T_0(0,0)$ will read:

$$t = \frac{r}{c} \qquad (1)$$

The events involved in the thought experiment described above are:

- $E_a(x_a, y_a, -\frac{r_a}{c}) = E_a(r_a\cos\theta_a, r_a\sin\theta_a, -\frac{r_a}{c})$ associated with the apparent position $S_a$,
- $E(x, y, 0) = E(r\cos\theta, r\sin\theta, 0)$ associated with the actual position S,
- $E_{0,1}(0,0,0)$ associated with the reception at O of the first light signal emitted from the apparent position $S_a$ and
- $E_{0,2}(0,0,\frac{r}{c})$ associated with the reception at O of the second light signal emitted from the actual position S.

The distance between $S_a$ and S is $V\frac{r_a}{c}$, the distance travelled by the source between its apparent and actual positions.

From the point of view of telemetry[2,3] $R_0$ represents a type of observer who collects information about the space-time coordinates of distant events from light signals that arrive at his location. The purpose of our paper is to present a space-time diagram that displays in true values the space-time coordinates of the events defined above.

Whereas Deissler[1] presents the coordinates of the apparent position as a function of those of the actual one, we present here the coordinates of the actual position as a function of the apparent position. Our results are simpler and more transparent.

**2. The space-time diagram displays in true values the space-time coordinates of events associated with the apparent and the actual positions of a point like source.**

Simple geometry applied to Figure 1 leads to:

$$x = x_a + V\frac{r_a}{c} = r_a(\cos\theta_a + \beta) \qquad (2)$$

$$y = r\sin\theta == r_a \sin\theta_a = d \qquad (3)$$

From (2) and (3) we obtain the following expression for the actual (dimensionless) position *x/d* of the source as a function of the polar angle $\theta_a$ that defines the apparent position:

$$\frac{x}{d} = \frac{\cos\theta_a + \beta}{\sin\theta_a} \qquad (4)$$

Figure 2 is a plot of $\frac{x}{d}$ as a function of $\theta_a$ for different values of $\beta$.

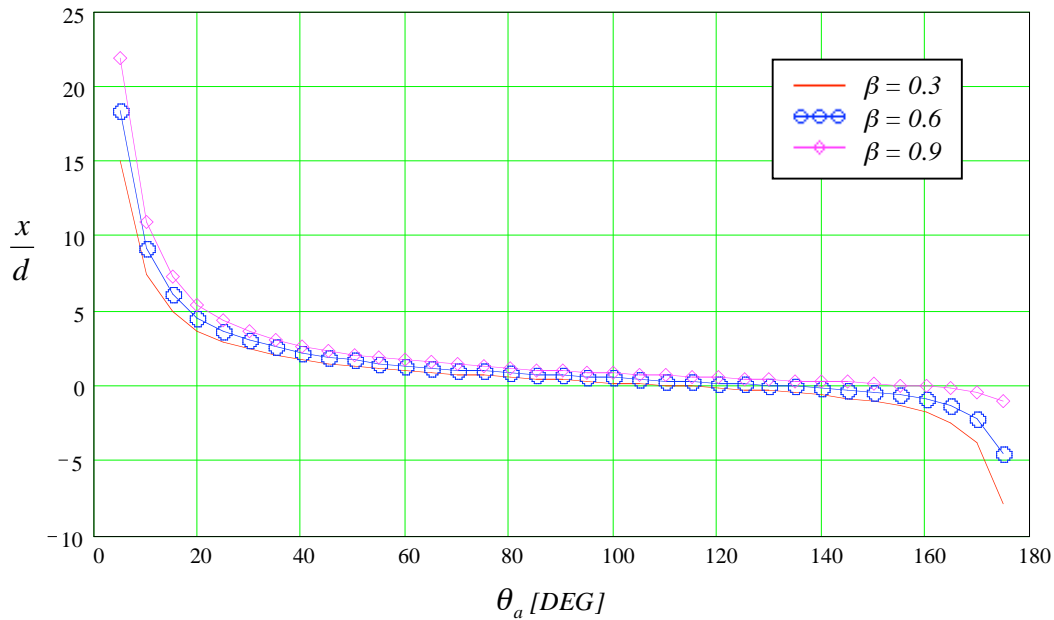

**Figure 2.** *A plot of $\frac{x}{d}$ as a function of the polar angle $\theta_a$ for three different values of the reduced velocity $\beta = \frac{V}{c}$ (0.3;0.6;0.9). For $0<\theta<\pi/2$ the source of light is "receding" whereas for $\pi/2 < \theta < \pi$ the source is "approaching" the OY axis.*

We obtain that the lengths of the position vectors that define the apparent and the actual positions are related by:

$$r = r_a\sqrt{1 + 2\beta\cos\theta_a + \beta^2} = \frac{d}{\sin\theta_a}\sqrt{1 + 2\beta\cos\theta_a + \beta^2}. \quad (5)$$

When S reaches the 2$^{nd}$ actual position clock $T_0$ reads:

$$t = \frac{r}{c} = \frac{r_a}{c}\sqrt{1 + 2\beta\cos\theta_a + \beta^2}. \quad (6)$$

The equations derived above enable us to construct a space-time diagram that displays in true values at well-defined scales the coordinates of the events associated with the apparent and actual positions. We present this diagram in Figure 3. Its axes coincide with the axes of the K(XOY) reference frame. The circle $C_a$ of radius $r_a$ having its centre at the origin O is the geometric locus of events $E_a$ and the circle $C$ defined by (4) is the geometric locus of events $E$. The invariance of the $y$ coordinates enables us to establish the position on the diagram of two corresponding actual and apparent positions $S$ and $S_a$ respectively.

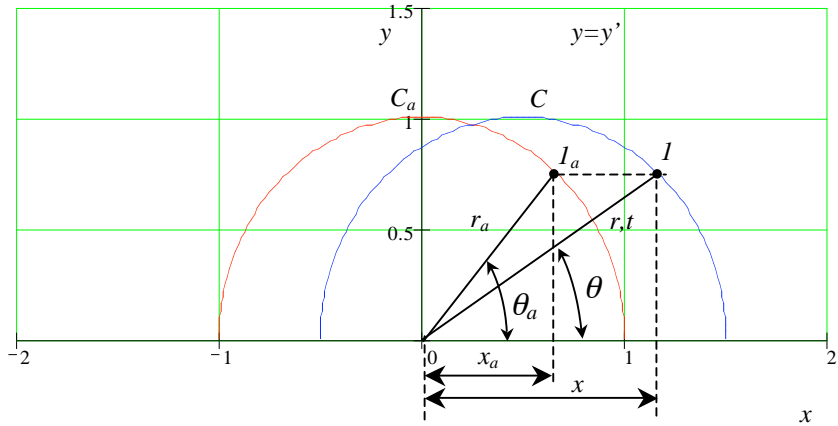

**Figure 3.** *The space-time diagram displays in true values the coordinates of the events associated with the apparent position $S_a$ and with the actual position S. It displays also the circle $C_a$ of radius $r_a=1$ having its centre at the origin O and the circle C described in parametric representation by (2) and (3) for $r_a=1$. The invariance of the $y(y')$ coordinates enables us to find out the location on diagram of the apparent and of its corresponding actual position. The diagram displays in true values the Cartesian and the polar coordinates of events $1_a$ and 1 as well as the readings $t=r/c$ of the clocks of K when observer $R_0(0,0)$ receives the light signal emitted from the actual position.*

Consider that $S_a$ represents one of the successive actual positions. Then $\frac{r_a}{c}$ represents the reading of clock $T_0$ when $R_0$ receives the light signal emitted from that position. Under such conditions, if at a given scale we have $r_a = c \cdot t_a$ then at the same scale $r = c \cdot t$.

Starting with (2) we allow for small changes in the value of variables. The result is:

$$dx = dx_a + \frac{V}{c} dr_a. \qquad (7)$$

Let *dt* be a small change in the readings of the clocks located at the different points of the K frame. By definition $\frac{dr_a}{dt} = V \cos\theta_a$ represents the radial component of the source speed at its apparent position, $\frac{dx_a}{dt} = V_a$ represents its actual speed and $\frac{dx}{dt} = V$ represents the OX component of the source speed. With the new notations (7) leads to:

$$\frac{V}{V_a} = \frac{1}{1 - \frac{V}{c}\cos\theta_a}. \qquad (8)$$

We present in Figure 4 a plot of $\frac{V}{V_a}$ as a function of $\theta_a$ for different values of $\beta = \frac{V}{c}$.

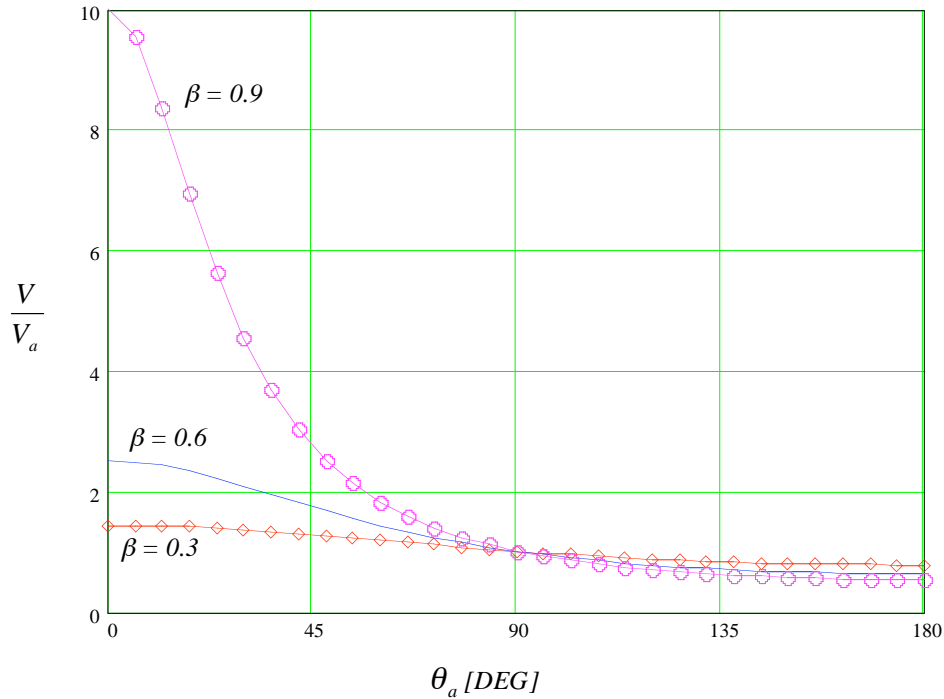

**Figure 4.** *A plot of V/V$_a$ as a function of the polar angle $\theta_a$ for three different values of the reduced velocity $\beta = V/c$.*

For $\theta = 0$ the source is receding and for $\theta = \pi$ the source is approaching the vertical axis. With $\beta$ defined above (8) becomes:

$$\frac{V}{V_a} = \frac{1}{1 \mp \beta} \quad (9)$$

recovering Deissler's result.[1] As we see, for $\theta_a = \pi/2$ the two velocities are equal to each other. In the receding conditions, $V > V_a$, whereas when the source is approaching $V < V_a$.

### 3. The space-time diagram at work

Consider that the geometric locus of the actual positions is a straight line parallel to the OY axis described in polar coordinates by:

$$r_a = \frac{a}{\cos\theta_a} \quad (10)$$

$a$ representing its distance to the OY axis. In accordance with (2) and (3) the geometric locus of the corresponding actual positions is described by the parametric equations:

$$x = a\left(1 + \frac{\beta}{\cos\theta_a}\right) \quad (11)$$

$$y = a \cdot tg\theta_a \quad (12)$$

The rules of handling the space-time diagram enable us to construct, point by point, the actual shape of (10), as shown in Figure 5.

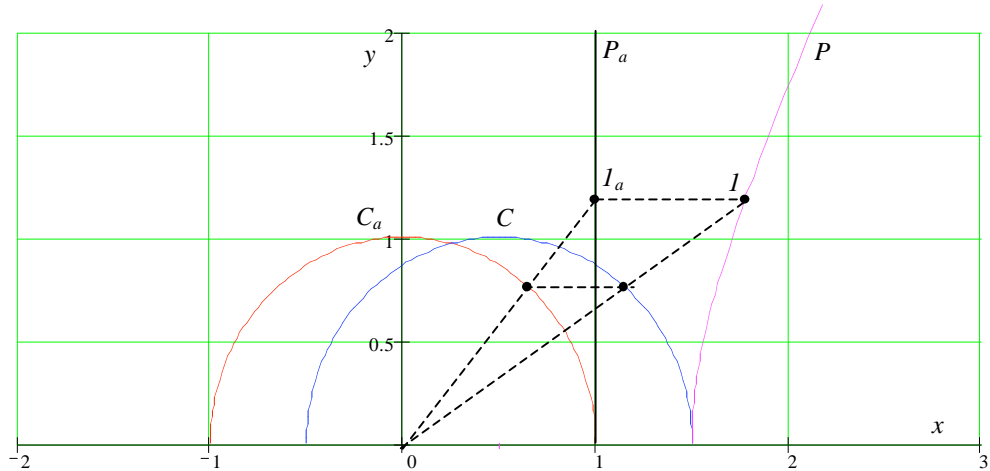

**Figure 5.** *The space-time diagram displays the basic circles $C_a$ and $C$. The vertical line $x=1$ ($P_a$) is the geometric locus of apparent positions of point like sources located on it, the curve (P) representing theirs actual position. We have constructed (P) point-by-point using the rules of handling our space-time diagram.*

## 4. Conclusions

We have constructed a space-time diagram that displays in true values the space-time coordinates of events associated with effects generated by the finite speed of light signals. Expressing the space coordinates of the actual positions as a function of those of the apparent positions we obtain simple and transparent results as compared with those presented by Deissler[1]. Our approach illustrates that the simple fact that light propagates with finite velocity relative to an inertial reference frame has anti common sense consequences like making a net distinction between "approaching" or "receding" positions of the light source.